\documentstyle[12pt]{article}
\newcommand {\be} {\begin{equation} \nonumber }
\newcommand {\bea} {\begin{eqnarray} \nonumber }
\newcommand {\ee} {\end{equation}}
\newcommand {\eea} {\end{eqnarray}}

\newcommand{\EC}{{\it E. Coli }}

\begin{document}        
\title{Complex Systems: a Physicist's Viewpoint}
\author{Giorgio Parisi\\
Dipartimento di Fisica, Universit\`a di Roma {\it La Sapienza },\\
 Piazzale Aldo Moro, Roma I-00185}
\maketitle
\section{Introduction}
In  recent years physicists have been deeply interested in studying 
the behavior of complex 
systems. The result of this effort has been a conceptual revolution, 
a paradigmatic shift that has 
far reaching consequences for the very definition of physics.

 In the nutshell, physics is an experimental science in which  
theoretical predictions are  
compared to experiments. It differs from other sciences of nature in 
the crucial role of 
mathematics:  physicists describe the phenomena  by using a 
mathematical 
model and their predictions are obtained by mathematical reasoning.

In this definition of physics the word {\it prediction} plays a 
crucial role. Naively one could think 
that the meaning of the word {\it prediction} is quite evident and 
no further explanation is needed. 
In reality its meaning has changed already in the past and it is 
still changing now. 

Let me recall  using modern language what  a prediction was at the 
time physics was born. We assume that 
the state of the system at the time $t$ is  represented by a a  
vector $X(t)$ with $N$ components. 

In the case of classical mechanics the state of a point-like 
particle is 
described by the three coordinates and by the three components of 
the velocity.  
The corresponding vector $X(t)$ has six components.  An experiment 
consists in determining or imposing the position 
and the velocity of the particle at a given time and measuring it a 
later time.  
It is crucial that (at least in principle) the experiment can be 
done many times 
by imposing the same initial conditions.  If the difference between 
the initial 
and the final time is the same, the system is always found in the 
same final 
state.  When repeated, the experiment always gives the same result.  
The 
theoretical task is to predict this unique, reproducible  result.

This can  be done by using the equations of motion which gives the 
time evolution of the  vector 
$X(t)$. In the case of classical mechanics, they have the form of  
Newton's second law  
\be
{dp\over dt} = F(x), \ \ \ \ \ {dx\over dt} ={p \over m}
 \ee
where $F(x)$ is the force. Once that the form of the force 
has been established it is a well posed 
mathematical problem to compute the final conditions as function of 
the initial one.

The same perspective applies not only to simple objects but also 
to much more complicated situations.  The description of the system 
may need 
 more than one position and one velocity.  The equations of motion, 
which 
are in general of the form
\be
{d X \over d t} = G(X(t)), \label{MOTION}
\ee
in principle allow the prediction of the final state for each 
initial state. The theory (which 
consists in assuming a particular form of the function $G(X)$) can 
be tested by  measuring $X$ at the 
initial time, predicting $X$ at the final time and verifying the 
prediction.

In reality, the mathematical difficulties may be substantial even if 
the system is apparently simple. 
To predict the evolution of a three body system (like Sun, Earth and 
Moon) is a rather difficult task 
and it can be done only with a lot a work. At the beginning of the 
nineteenth century Laplace  was 
writing that an infinitely intelligent mathematician would be able 
to predict with certitude the 
future by  observing the state of the universe now and using his 
knowledge of the laws of motion.

In this classic, deterministic framework the word prediction has a 
clear meaning. Unfortunately
 we are not 
infinitely intelligent mathematicians and besides an 
infinitely able experimentalist is needed to measure the state of 
the 
system with extremely high accuracy. The Newton-Laplace point of 
view could be applied only to a 
restricted class of phenomena. In order to consider other phenomena 
which could not be studied from 
this point of view, it was necessary to change the general {\it 
philosophy}, by introducing 
probabilistic concepts and probabilistic predictions. 

 There have been three revolutions in physics which all went in this 
direction \cite{DERAMO}
  and as a consequence 
they have changed  the meaning of the word {\it prediction}. They 
are:
\begin{itemize}

\item (1) The introduction of statistical mechanics and of the first 
probabilistic reasoning by 
Maxwell, Boltzmann and Gibbs in the second half of the last century.

\item (2) The discovery of quantum mechanics at the beginning of 
this century.

\item (3) The study of complex systems and the related techniques 
that have been developed 
in these 
last years.

\end{itemize}

As an effect of these revolutions, the word {\it prediction} 
acquired a weaker 
meaning.  Predictions in the context of the new paradigm are not 
acceptable with 
the old one (and sometimes the supporters of the old point of view 
try to deny 
to them a scientific validity).  The positive consequence of the 
process is that 
the scope of physics becomes much larger and the constructions of 
physics find 
many more applications.

 My aim is to present point (3). However, it is better to   start by 
discussing point (1) in order to 
understand in which way the study of complex systems forces us to 
use the word probability in a 
wider context. For our purpose we can neglect point (2), because 
extends the 
concept of  
probability in a different direction \footnote{In quantum mechanics 
we have to deal with the fact that also in principle the results of 
a single experiment are not reproducible.}.

\section{Introducing Probability in Physics}

At the time of Boltzmann \cite{BOL} and Gibbs, the main motivation 
for leaving 
the Newton-Laplace point of view was not that it was wrong, but that 
it was 
useless in many cases.  Let us consider the following experiments 
which 
apparently can be repeated many times with the same result: 
\begin{itemize}
\item We bring water to 100 centigrade: it boils.
\item We bring water to 0 centigrade: it freezes.
\end{itemize}

When we study this phenomenon from a microscopic  point of 
view,  we face a serious problem. 
The experiment should be done by measuring the positions and the 
velocities of all the billions of 
billions of atoms and we should later use this information to 
compute the 
trajectories of all the 
atoms. This difficulty was bypassed by discovering that for a system 
composed by many many particles, 
practically all the initial conditions (at fixed total energy) lead 
to the same macroscopic behavior of the system. The task of 
measuring everything 
is not only impossible, it is also useless. If we neglect the 
possibility of some very special 
initial configuration - which happens with extremely low probability 
- the system always behaves in 
the same way. 

The final predictions are the following:
\begin{itemize}
\item
Water practically always boils at 100 degrees. The probability is so 
high that it is extremely likely  
that this kind of predictions has never failed in the whole history 
of the universe.
\item
If we measure the velocity of a single molecule of water we cannot 
predict anything precise. However 
we can compute the probability that the  molecule has a velocity $v$
($P(v)\propto \exp(-A 
v^2)$ where $A=m/(2kT)$). Therefore if we measure the velocity of a 
molecule $M$ times the observed distribution 
of velocity should approach the theoretical curve when $M$ goes to 
infinity.
 \end{itemize}

In general, we could say that for these large systems there are two 
kinds of quantities:
\begin{itemize}
\item
Some quantities  can be predicted with certitude and have always the 
same value inside a small interval (usually of order $N^{-1/2}$ for 
a system with $N$ degrees of freedom).
\item
Other quantities do not take always the same value. In this case a 
probabilistic prediction is possible.
\end{itemize}

\section{Deterministic Chaos}

In the previous discussion, the main motivation for introducing 
probability was 
the large number of particles present.  Only in the second half of 
this century 
 was it realized that there are many systems with a small number of 
particles, or 
more generally with a small number of degrees of freedom, for which 
it is also 
necessary to use probabilistic arguments \cite{RUE}.

 In these systems (which sometimes are called chaotic) the 
trajectory cannot be predicted for large
times, because it is too sensitive to the initial conditions. The 
difference in the position of the
trajectories of two systems which have a very similar initial 
conditions increases exponentially
with time. In other words, even a very small incertitude on the 
initial condition leads to a total
loss of knowledge after a characteristic time $\tau$. More precisely
if 
\be
X_1(0)-X_0(0)\propto \epsilon
\ee
the difference $X_1(t)-X_0(t)$ grows as function of the time $t$ as
$
\epsilon \exp (t/\tau)
$
until it reaches a value of order 1.

Deterministic predictions are possible here on a time scale smaller 
than $\tau$, but they become
impossible at later times because we cannot measure the initial 
conditions with infinite precision.

A very simple example is provided by two (or more) balls on a 
billiard table 
without friction.  For generic choice of the initial condition the 
two balls 
will collide from time to time, and after each collision it becomes 
more 
difficult to predict the position of the balls.

Fortunately, in many cases the probability distribution at large 
times for finding 
the
system in a given configuration is independent on the initial
conditions and it can be predicted. So here also  we can compute 
only the probability distribution of
some variables, not the exact evolution.

The new framework given by statistical mechanics should now be 
clear.  We have a 
system, we know the equation of motion.  In principle, an exact 
knowledge of the 
initial conditions allow us to compute the exact evolution of the 
systems.  In 
reality this task may be impossible because we do not know the 
initial 
conditions with sufficiently  high accuracy.  It is also possible 
that the 
computation is terribly complicated.  We can however make progress 
and get 
much more insight if we restrict ourselves to the task of predicting  
the probability distribution of the 
system at large times (in short its {\it behavior}).
Given the  equation of motion $G$ (see eq. \ref{MOTION}) we define
the probability $P_G(X)$ \cite{RUE} as
\be
\int dX f(X) P_G(X)= \lim_{T\to \infty}{\int_0^T f(X(t))dt \over 
T},
\ee
where $f(x)$ is an arbitrary test function. For ergodic systems
the probability $P_G(X)$ does not depend on the choice of the 
initial point for almost all the choice.

The task of statistical mechanics consists in the computation for 
given $G$ of 
the probability $P_G(X)$  and of its properties.
\section{Complex Systems}

It was recently realized that this approach described in the 
previous section  (i.e.  to predict the 
behavior of the system from the knowledge of the equations of 
motion) fails in 
the case of complex systems for reasons that are very similar 
(albeit in a 
different contest) to those which led to the abandonment of the 
Newton-Laplace view-point.

There are many possible definitions of a complex system.  I will use 
the 
following one.  {\it A system is complex if its behavior crucially 
depends on 
the details of the system.} This dependence is often very difficult 
to 
understand \footnote{ 
A single complex system may also display different types of behavior 
and a small 
perturbation is enough for switch from one behavior to another.  For 
example an 
animal may: sleep, dream, run, hunt, eat, play...}.  In other words, the behavior of the system may be 
extremely 
sensitive to the form of the equations of motion and a small 
variation in the 
equations of motion leads to a large variation in the behavior of 
the system. More precisely 
for two systems with $N$ degrees of freedom with equations of motion
$G_0$ and $G_1$, where
\be
G_1(X)-G_0(X)=O(\epsilon),
\ee
we could have that for small $\epsilon$ but large
$\epsilon N$,
\be
P_{G_1}(X)- P_{G_0}(X) =O(1).
\ee
In this happens it is practically impossible to compute the 
asymtotic probabily as function of $G$.

My aim is to show that although a system is complex, it is still 
possible to get 
predictions for its behavior, but these predictions will be now of a 
probabilistic nature.  I will illustrate this point in a particular 
example.  
The prototype of a complex system in physics is spin glass 
\cite{MPV}.  I will 
choose here something more familiar to most people, a large protein.

A protein is composed by a long chain of a few hundred amino-acids 
and its 
chemical composition is specified by the sequence of aminoacids 
(primary 
structure).  In physiological situations, normally a protein folds 
in an unique 
way (tertiary structure).  During the folding, the protein goes to 
the 
configuration of minimal energy (more precisely, to the 
configuration of minimal 
free energy).  However, there are many quite different 
configurations of the 
protein which have an energy near the minimal one.  There are 
proteins, which 
are called allosteric, which have two configurations which 
practically the same 
energy). Sometimes, if we perturb the protein by a small amount, 
e.g.  changing the pH or changing only one aminoacid in the chain, 
the folding 
changes dramatically.  We can also select one of the two different 
configuration by the binding of a ligand.

The variations in the shape of mioglobulin in different conditions 
are crucial 
for its respiratory functions.  More generally, the possibility of 
switching 
from one configuration to another is one of the fundamental 
mechanisms which 
allows proteins to work as enzymes.  For example the change of 
configuration of 
allosteric proteins it is used for doing useful work, e.g.  for 
contracting muscular fibers.

The existence of allosteric proteins implies that a small variation 
in the form of the potential among 
the components of the protein changes the folding of the protein 
dramatically.  
It is thus evident that a small error in the form of the interaction 
among the 
different components of the protein or of the interaction with the 
solvent would 
lead to quite wrong results.

Natural proteins are on the borderline of what can  actually be 
understood; we have not yet
computed their folding properties, but this does not seem completely 
impossible, especially for
the smallest ones. However, to compute the folding properties of 
much large proteins seems to be a
hopeless task.

If an approximate knowledge of the equations of motion does not 
allow us to compute the behavior of
a complex system, we can give up this task and restrict ourselves to 
compute the probability  of a
behavior when the equation of motions are chosen randomly inside a 
given class.

In other words we can study the same problem in two different ways: 
\begin{itemize} 
\item In the old approach we consider a given protein (e.g.  mio\-
globulin) and we 
compute the form of the two foldings with minimal energy and their 
energy 
difference as a function of the chemical properties of the solvent.
\item
In the new approach we consider a class of proteins (to which 
mioglobullin belogns) and we compute the probability
distribution of the energy difference among the two foldings with
minimal energy \cite{SG}.
\end{itemize}

Of course, this second computation will not tell us the properties 
of 
mioglobulin.  It will give us predictions that cannot be tested  on 
a single 
protein (in the same way the probability distribution of the 
velocities cannot 
be tested by measuring only one velocity); we need to repeat the 
experiments on 
many different proteins.

Taking the new point of view, we often gain insight. The behavior of 
a given
complex system may be obtained only after a very long computation on 
a computer and often we cannot
understand the deep reasons of the final result. The study of the 
probability distribution of the
behavior can sometimes be done analytically and we can follow all 
the steps that lead to the
conclusions.

The advantage of the second point of view is that its results may be 
easily 
generalized to other systems.  For example long RNA molecules also 
fold in a 
characteristic way, in the same way that proteins do.  A general 
explanation which shows that some 
characteristic of folding are shared by all polymers composed by 
different monomers may be applied 
to RNA as well as to protein

There are simple models (e.g. spin glasses \cite {MPV}) in which 
this program works. The computation of the
behavior of a given system is extremely difficult and sensitive to 
the minimal detail and can be
done only with lengthy computer simulations. The analytic 
computation of the probability
distribution of the behavior for a generic system can be done 
analytically and these probabilistic
predictions have been successfully verified.

The extension of this program to other systems leads to two 
different types of difficulties.
\begin{itemize}
\item The generalization of the methods which allow the computation 
of the probability distribution
of the behavior to  models which are not as simple as spin glasses, 
is technically difficult and 
progress in this direction are rather slow.
\item
This point of view is quite different from the traditional one 
(mathematically it corresponds to the use of {\sl imprecise 
probabilities}\cite{IP}).  The predictions 
are not made for the properties of a given system, but for the 
probability 
distribution of those properties which change with the system.  We 
are not 
enough familiar with this new point of view to appreciate all its 
potentialities.
 \end{itemize}

As usual a change in the paradigm leads to a change in the questions 
posed. We do not have
to ask how a particular system behaves. We have to ask which are the 
general features of the
behavior of a system belonging to a given class.

In spite of these difficulties, this new approach seem to be 
absolutely necessary to get some
understanding in really complex systems. We have seen that the 
arguments are very similar to those
described before. In the two cases they can be summarized as follows
 \begin{itemize}
\item (1) The extreme sensitivity of the trajectory on the initial 
conditions forces us to study the
probability distribution of the systems at large times. This must be 
done in spite of the fact that
an exact knowledge of the initial condition determines the 
trajectory at all times. The behavior
of the system can be computed from the equations of motion.
\item (2) The extreme sensitivity of the behavior on the detailed 
form of the 
equation of motion forces us to study the probability distribution 
of the 
behavior of the system.  This must be done in spite of the fact that 
an exact 
knowledge of the initial conditions determines the behavior.  The 
probability 
distribution of the behavior of the system can be computed from the 
probability 
distribution of the equations of motion. In other terms, given a 
probability distribution $\rho(G)$ we can compute the probabillity
${\cal P}[P_G(X)]$, which of course depends on $\rho$.

\end{itemize}

The advantage of this new point of view is to expand the range of 
application of
physical reasoning. We shall see now how it puts the relations 
between physics and biology in a new
perspective.

\section{Possible Biological Applications}
 In biology problem we face a very difficult of how to use the 
immense amount of information we
gather at the molecular level in order to understand the behavior of 
the whole organism.

 Let us consider an example. One of the simplest organisms is \EC. 
Its genome (which is known) codes for about 3000
different proteins which interacts among themselves. Some proteins 
promote the production of other
proteins, while other proteins have a suppressor effect. 

In principle we can gather the information
on  the properties of the proteins, on the interaction among the 
proteins (for
simplicity let us neglect  all the other chemical elements of the 
cell). At the end we can use this
information to construct a model of the living cell. The model may 
lead to a system of few thousand
(or more) coupled differential equations which can be studied by a 
lengthy computations on a 
computer.

If the information is accurate enough, the model will describe a 
living cell.  
However we know that in real life  many mutations are lethal, and 
therefore 
it is quite likely that imprecision in the form of the interactions 
among 
proteins will lead to a non-living cell or to a living cells with a 
quite 
different behavior.  It is also clear that such a gigantic 
computation (although 
welcome) will not capture the essence of life.  Indeed it should be 
repeated 
nearly from scratch if instead of \EC   we consider a different 
organism.

The previous discussion suggests that we could give up the aim of 
deriving the properties of a
given organism from its chemical components, and ask different 
questions, e.g. 
what are the general
properties of a living organism and how they do change from organism 
to organism.

 An example of this
last of question has been studied by Kaufmann \cite{K}. There is a 
dependence of the number 
different cell
types on the number of genes  in all
kind  of living organisms. 
\be
\# \  cell\  types \propto (\# \  genes)^{1/2}
\ee

As Kaufmann points out, this fact must have a general explanation 
and he
proposes one based on mathematical considerations of evolution.

An other empirical law, which calls for a general  explenation  is
\be
\# \ species \ in \ a  \ island \propto (\# \  genes)^{1/4}
\ee

In these two cases we can hope to compute the exponents. However in many 
other cases this point of view is so different form the traditional 
one, that it is very 
hard
to find the right questions and also to answer to them.

A field in which this approach seems to be mandatory is the study of 
the origin of life. Here it is
crucial to understand which are the enzymatic capabilities of the 
prebiotic material (maybe long 
randomly
assembled RNA chains) before evolution starts. We would like to know 
which of the properties of
these long chains had to be selected and which ones where already 
present before selection.

This approach is quite different from
a redutionistic one, as far as it puts the stress on the behavior of 
the whole system.
For the moment the largest impact of these ideas in biology has been 
in the field of neural
networks, but in the other fields the progress is quite slow. We 
need to sharpen our theoretical
tools in order to be able to predict the typical behavior for  
different classes of systems. We are
seeing now only the beginning of these investigations in 
mathematical physics. 

When the physical instruments become more robust and our theoretical 
command 
increases, the interaction with biology will become more easy.  I am 
convinced 
that in the next century a much more deep understanding of life will 
come from 
this approach.

\end{document}